\documentclass[prb,twocolumn,draft,amsmath,showpacs]{revtex4}
\usepackage{graphics}

\begin{document}

\input epsf

\title {Magnetothermal transport in the spin-$1/2$ chains of copper pyrazine dinitrate}

\author{A. V. Sologubenko,$^1$ K. Berggold,$^1$ T. Lorenz,$^1$ A. Rosch,$^2$  E. Shimshoni,$^3$ 
M. D. Phillips,$^4$ and M. M. Turnbull$^4$}

\affiliation{$^1$II. Physikalisches Institut, Universit\"{a}t zu K\"{o}ln, 50937 K\"{o}ln, Germany}
\affiliation{$^2$Institut f\"{u}r Theoretische Physik, Universit\"{a}t zu
  K\"{o}ln, 50937 K\"{o}ln, Germany}
\affiliation{$^3$Department of Mathematics--Physics, University of Haifa at Oranim, Tivon 36006, Israel}
\affiliation{$^4$Carlson School of Chemistry and Department of Physics, Clark University,
Worcester, MA 01610, USA}
\date{\today}

\begin{abstract}
We present experiments on the thermal transport in the   spin-$1/2$ chain compound copper pyrazine dinitrate Cu(C$_4$H$_4$N$_2$)(NO$_3$)$_2$. The heat conductivity  shows a surprisingly strong dependence on the applied magnetic field $B$, characterized at low temperatures by two main features.
The first one appearing at low $B$  is a characteristic dip  located at $\mu_B B \sim k_B T$, that may arise from Umklapp scattering.  
The second one is  a plateau-like feature  in the quantum critical regime, $\mu_B |B-B_c|<k_BT$, where $B_c$ is the saturation field at $T=0$.
The latter feature clearly points towards a momentum and field independent mean free path of the spin excitations, contrary to theoretical expectations.

\end{abstract}
\pacs{
75.40.Gb 
66.70.+f, 
75.47.-m 
}
\maketitle

In the last decade,  considerable progress has been achieved in
theoretical studies of thermal transport in one-dimensional (1D)
quantum spin systems (for a recent review see
Ref.~\onlinecite{Zotos05_Rev}). One of the most important model systems is
the Heisenberg spin $S=1/2$ chain with isotropic antiferromagnetic
interactions, described by
\begin{eqnarray}
\label{eHamiltonian}
    H = J \sum_{i}^{N} {\bf S} ^{i} {\bf S}^{i+1}- g \mu_B B
    \sum_{i}^{N} S_z^i,
\end{eqnarray}
where $J$ is the intrachain nearest-neighbor exchange and $g
\mu_B$ the magnetic moment.
Recently, a number of non-trivial effects were predicted for the spin
thermal conductivity ($\kappa_s$) of this system in external magnetic
fields $B$.\cite{Louis03,Shimshoni03,Sakai05_MTE,Sakai03_XXZ,HeidrichMeisner05,Savin05}
Of particular interest is the behavior close to the saturation field
$B_c=2J/g\mu_{B}$, which defines a quantum critical point.\cite{SachdevBook}
 Experimental
information on $\kappa_s(B)$ is, however, missing because most
of the known realizations of the Hamiltonian
(\ref{eHamiltonian}) have large values of the intrachain exchange
constant $J/k_B \sim 100 - 1000$~K, that severely limits the region of
the phase diagram accessible using standard laboratory equipment.  The
existing results are limited to studies of the phonon thermal
conductivity $\kappa_{\rm ph}(B,T)$ in CuGeO$_3$ and Yb$_4$As$_3$,\cite{Ando98,Takeya01,Hofmann02,Koeppen99} and do not address the
behavior of $\kappa_s(B,T)$.

Copper pyrazine dinitrate Cu(C$_4$H$_4$N$_2$)(NO$_3$)$_2$ (CuPzN)
appears to be an ideal compound for the thermal conductivity
experiments in magnetic field, as $J/k_B=10.3$~K
 and therefore $B_c=15.0$~T,\cite{Mennenga84,Hammar99} which is easily
accessible experimentally.  CuPzN has an orthorhombic structure with
lattice constants $a=6.712$~\AA, $b=5.142$~\AA, and $c=11.73$~\AA ~at
room temperature.\cite{Santoro70} The chains of Cu$^{2+}$ spins
$S=1/2$ run along the $a$ axis.  Inelastic neutron scattering,
magnetization, and specific heat measurements have confirmed that
CuPzN is very well described by the model of Eq.~(\ref{eHamiltonian}).\cite{Mennenga84,Hammar99,Stone03}  The 1D nature of the spin
interaction is reflected by a very low ordering temperature,
$T_N=0.107$~K, and therefore the ratio of interchain ($J'$)
to intrachain ($J$) couplings is estimated to be tiny, $|J'/J| \approx 4.4 \times
10^{-3}$.\cite{Lancaster06}

In this paper, we report measurements of the thermal conductivity
$\kappa(B,T)$ of CuPzN in the temperature region between 0.37 and 10~K
and in magnetic fields up to 17~T. 
The crystals of CuPzN were grown from water solution of pyrazine and
Cu nitrate via slow evaporation.  The crystals have right-prism shapes
with the $a$ axis of length 10~mm directed along the height of the prism. The dimensions
perpendicular to the $a$ axis are typically of the order of
0.4$\times$0.7~mm$^2$.  Thermal conductivity was measured by a
standard steady-state heat-flow technique, where the temperature
difference was produced by a heater attached to one
end of the sample and monitored by a matched pair of RuO$_2$
thermometers. The temperature difference between the thermometers was
of the order of 1\% of the mean temperature $T$.  
In one sample the heat
flux was oriented along the $a$ axis to measure the thermal
conductivity parallel to the chains ($\kappa^{\parallel}$).
On another sample,  the thermal transport perpendicular to the chains ($\kappa^{\perp}$) was measured.
Magnetic fields were oriented  parallel to the $a$~axis.

\begin{figure}[t]
   \begin{center}
    \leavevmode
    \epsfxsize=0.8\columnwidth \epsfbox {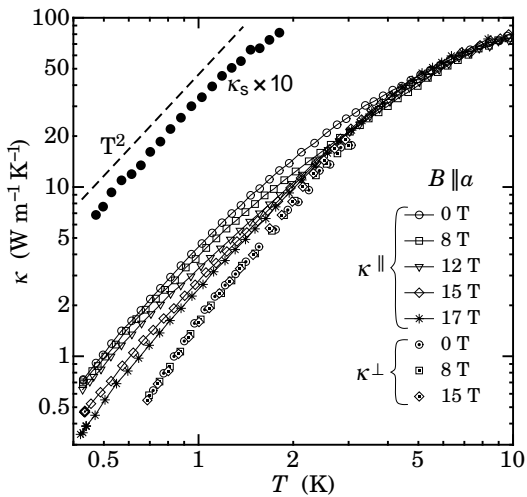}
     \caption{
     The thermal conductivity of CuPzN parallel and perpendicular to the spin chains as 
     a function of temperature in several constant magnetic fields. 
The solid circles correspond to the calculated  zero-field spin thermal conductivity along the chains (see text).
    }
\label{KT}
\end{center}
\end{figure}

In Fig.~\ref{KT}, we show the temperature dependence of
$\kappa^{\parallel}$ and $\kappa^{\perp}$ in constant magnetic fields.
The important observation is that $B$ strongly influences
$\kappa^{\parallel}$, but no significant changes of
$\kappa^{\perp}$ with field are observed. 
In CuPzN, the magnetic interaction between the spin chains is extremely
weak. Therefore one expects that the heat current perpendicular to the
chains is of purely phononic origin, $\kappa^{\perp} \approx \kappa^{\perp}_{\rm ph}$, consistent
with the observed absence of a significant field dependence. Along the
chain direction, $\kappa^{\parallel} =\kappa^{0\parallel}_{\rm
  ph}+\kappa_m$, one can separate the field-independent
phononic contribution, $\kappa^{0\parallel}_{\rm ph}(T)$, unrelated to the presence of the spin chains, from a $B$ dependent magnetic
contribution, $\kappa_m(B,T)$. Three terms contribute to
$\kappa_m=\Delta \kappa^{\parallel}_{\rm ph}+ \kappa_s +\kappa_{sp}$.
First, the scattering by spin excitations affects the lifetime of
phonons and gives rise to a decrease $\Delta \kappa^{\parallel}_{\rm
  ph}$ of the phononic thermal conductivity. 
Second, there is a spin contribution $\kappa_s$ to the thermal transport. Finally, a spin-phonon
cross term $\kappa_{sp}$ describes the `drag' of spin heat
currents by phononic heat currents (and vice versa).  
For  constant $T$, the difference between the total measured
$\kappa^{\parallel}(B)$ and its zero-field value
$\kappa^{\parallel}(B=0)$ allows one to extract the field
dependence of the magnetic heat conductivity
$\kappa_m(B,T)-\kappa_m(0,T)$.

\begin{figure}[t]
 \begin{center}
  \leavevmode
  \epsfxsize=1\columnwidth \epsfbox {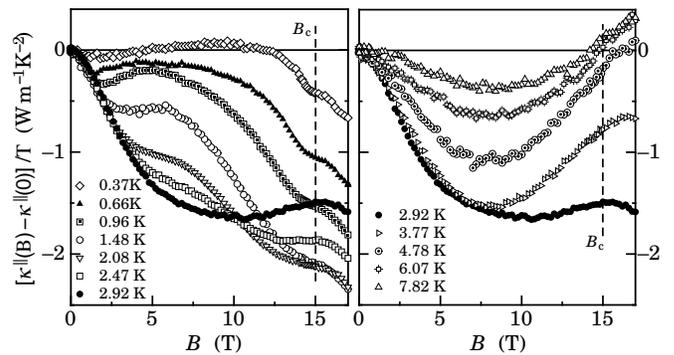}
   \caption{
   Thermal conductivity parallel to the chains of CuPzN as a function of  $B \parallel a$ at several fixed temperatures.    }
\label{KH}
\end{center}
\end{figure}

\begin{figure}[t]
 \begin{center}
  \leavevmode
  \epsfxsize=0.6\columnwidth \epsfbox {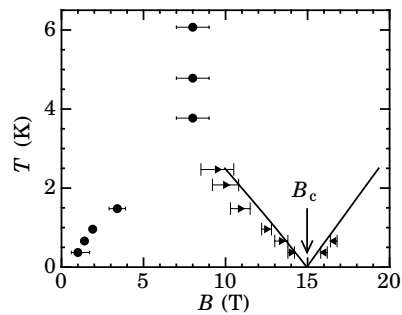}
  \caption{A ($B$,$T$) diagram of the characteristic features of the thermal
    conductivity of CuPzN. The circles correspond to the minima of
    $\kappa_m(B)$. The triangles correspond to the inflection points
    of $\kappa_m(B)$ in the vicinity of $B_c$. The solid lines are the
    positions of the $\kappa_m(B)$ inflection points calculated for
    a constant mean free path (see text).  }
\label{PhDia}
\end{center}
\end{figure}

In Fig.~\ref{KH}, we plot $[\kappa^{\parallel}(B)-
\kappa^{\parallel}(0)]/T \approx [\kappa_m(B)-\kappa_m(0)]/T$ at
several constant temperatures.  Depending on $T$, one can
distinguish two types of behavior. The crossover between these two
types occurs at about 2.5 -- 3~K.  At low temperatures, $T \ll J/k_B$,
$\kappa_m (B)$ has two main features: a decrease with a minimum at low
fields and another decrease at higher fields. 
On top of the high-field decrease, there is a plateau-like feature in the vicinity of the
critical field $B_c$. 
At high temperatures, $\kappa_m(B)$
has a simpler shape with a single minimum close to $B_c/2$.
The circles in Fig.~\ref{PhDia} show the positions of the $\kappa_m(B)$ minima. 
In the same figure the triangles, determined as the inflection points of  $\kappa_m(B)$ curves, characterize the size of the plateau-like regions in the vicinity of $B_c$.

For the interpretation of these data, we first note that 
a spin gap $=g \mu_B (B-B_c)$ opens when $B$ exceeds  $B_c$.\cite{Mueller81}
Therefore, for $g \mu_B (B-B_c) \gg k_B T$, both heat transport by spin excitations and phonon scattering by spin excitations should vanish.
In our experiment, $\kappa_m(B)$ decreases with  increasing  $B > B_c$ for low $T<2.5$~K. 
Thus, the observed field
dependence of the thermal conductivity arises dominantly  from thermal transport in
the spin system and not from  $\Delta \kappa^\|_{\rm ph}$. 
The overall decrease of $\kappa_m$ towards lower $T$, see
Fig.~\ref{KT}, suggests that impurity scattering dominates spin-phonon scattering. Therefore, spin-phonon drag
terms should not be important. This allows us to associate $\kappa_m$ with
the spin thermal conductivity $\kappa_s$.
We model $\kappa_s$ using a combination of a
relaxation time approximation and a mean-field theory (MFT)
for the spin chain relying on mapping Eq.~(\ref{eHamiltonian}) onto a system of
interacting spinless fermions via the Jordan-Wigner transformation.\cite{Jordan28}  The total weight of the frequency-dependent spin thermal conductivity,
$\int \kappa(\omega) d\omega$, in Heisenberg chains
calculated within the MFT approach has been shown to be in reasonable
agreement with exact results of the Bethe ansatz calculations.\cite{HeidrichMeisner05,HeidrichMeisner05PhD}  The fermions occupy a
cosine band which, at $B=0$, is half-filled with the
chemical potential at $k_F = \pm \pi/2a$. An external magnetic field
shifts the chemical potential and changes the band width.  The
dispersion of the fermions is
\begin{equation}\label{eMFTdisp}
\varepsilon_k = - J (1 + 2 \Omega) \cos(ka) - g \mu_B B + 2 J m, 
\end{equation}
 where $k$ is the wave vector and  $a$ is the distance between neighboring spins. 
The parameter $\Omega$ and the average local magnetization $m$ are
determined self-consistently from $\Omega =  \frac{a}{\pi}\int_{0}^{\pi/a}{\cos(ka) f_k dk }$
and $
m =  -\frac{1}{2} + \frac{a}{\pi} \int_{0}^{\pi/a}{f_k dk }$ 
where $f_k = (\exp(\varepsilon_k/k_B T)+1)^{-1}$ is the Fermi distribution function. 
The spin thermal conductivity is given by 
\begin{eqnarray}\label{eMFTKappa}
\kappa_s &=&  \frac{Na}{\pi} L_2,\\
 L_n &=&  \int_{0}^{\pi/a}{\frac{df_k}{dT} \varepsilon_k^{n-1} v_k^2 \tau_k   dk}, 
\end{eqnarray}
$N$ is the number of spins per unit volume, $v_k = d\varepsilon_k/dk$
is the velocity and $\tau_k$ is the relaxation time related to the mean
free path $l_k$ as $\tau_k = l_k/v_k$.

\begin{figure}[t]
 \begin{center}
  \leavevmode
  \epsfxsize=1\columnwidth \epsfbox {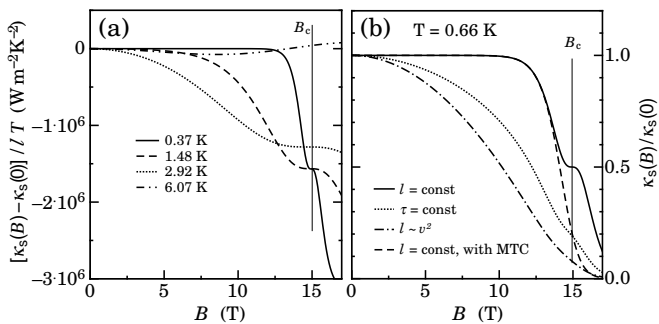}
   \caption{  (a) Spin thermal conductivity 
 as a function of  $B $ for several $T$, calculated for CuPzN within a relaxation time
     approximation assuming a $B$- and momentum-independent mean free path $l$.
   (b) Spin thermal conductivity at $T=0.66$~K, calculated under various
assumptions for the mean free path (see legend), and taking into
account magnetothermal  corrections (dashed line). Only the model with
constant $l$ is consistent with the data.
   }
\label{KHcalc}
\end{center}
\end{figure}

As discussed above, scattering by defects is expected to be the most important source of
extrinsic scattering at low $T$.
Assuming for the moment a mean free path,
$l_k(B,T)=l(T)$, which is independent of both magnetic field and
momentum,
 we can calculate $\kappa_s l^{-1}$ from Eq.~(\ref{eMFTKappa})
as a function of $B$ and $T$ without
free parameters.
 For our calculations, we used the experimental
values of the $g$-factor parallel to the $a$-axis $g=2.05$ from
Ref.~\onlinecite{McGregor76} and $J/k_B=10.3$~K from Ref.~\onlinecite{Hammar99}.
The resulting $[\kappa_s(B)- \kappa_s(0)] /lT$ is shown  in
Fig.~\ref{KHcalc}(a) for several $T$. 
The calculations obviously give correct qualitative account for the
experimentally observed behavior at
intermediate and high fields for all $T$. It reproduces the crossover
from the low-$T$ to the high-$T$ behavior around $3$\,K and predicts
the low-$T$ plateau-like features at $B_c$ with
the correct width. In Fig.~\ref{PhDia}, the calculated inflection points of $\kappa_s(B)$ (solid lines) are in good agreement with the experiment.  
 Close to the saturation field, i.e. at the
quantum critical point, one finds the following scaling relation
\begin{eqnarray} \label{scaling}
\frac{\kappa_s}{T l}&\approx & \frac{N a k_B^2}{\hbar} \,
f\!\left(\frac{g \mu_B  (B-B_c)}{k_B T}\right),\\
f(x)&=& \int_0^{\infty} \frac{(x-y)^2}{ 4 \pi \cosh[(x-y)/2]^2} d y \nonumber\\
&\approx & \frac{\pi}{6} \left\{ 
\begin{array}{ll}
2-3 e^{x} x^2/\pi^2&,  x\ll -1\\
1- x^3/(2 \pi^2) &, |x| \ll 1  \\
 e^{-x}( x^2+2 x+2)6/\pi^2 &, x\gg 1
\end{array}
\right.\nonumber
\end{eqnarray}
where  $f(x)$ is a dimensionless scaling function independent of the precise dispersion of the spin excitations.  

The comparison of the theoretical curves for various assumptions on
the momentum dependence of the scattering rate, Fig.~\ref{KHcalc}(b),
with our experimental results, Fig.~\ref{KH}, clearly supports a mean
free path which depends on $T$ but {\em not} on momentum and
magnetic field at least close to the quantum critical point. Note that
close to $B_c$, the mapping of the spin chain to free fermions becomes
exact and one may therefore hope that a simple Boltzmann description
becomes asymptotically exact (if strong localization by disorder can be
neglected). Here one has to emphasize, that the observation of
a constant mean free path is highly surprising in a
one-dimensional system.  In the
limit of weak disorder (small potential strength compared to $T$ or
$|B-B_c|$) the golden-rule scattering rate $1/\tau_k$ is proportional to the
density of states, $1/\tau_k \propto 1/v_k$, where $v_k$ is the
velocity of the spin excitations, and therefore one gets $l_k \propto
v_k^2$. Luttinger liquid corrections to this formula are
expected to vanish close to the quantum phase transition when the density of spin
excitations is small.  
In the opposite limit of strong impurities, the
spin excitations have to tunnel through the potential and the
conductivity is proportional to the transmission rate,  $T_{k} \propto v_k^2$, implying that 
$l \propto T_{k} \propto  v_k^2$, as in the limit of weak disorder. 
However,  the heat conductivity, calculated assuming 
$l_k \propto v_k^2$, is featureless at $B\approx B_c$, in clear disagreement with the experiment,
see Fig.~\ref{KHcalc}(b).

What are possible mechanisms which can lead to  a constant mean
free path? In the absence of inelastic scattering,  interference
effects lead to localization of the spin excitations but neither the observed
$T$ dependence of $\kappa$, see Fig.~\ref{KT}, nor the
constant mean free path point towards the importance of localization
effects. Nevertheless, rare defects may cut the spin chains into
separate pieces of finite length. In such a situation, the effective mean free path is determined
either by scattering from one spin chain to the next or -- more likely
-- by coupling heat currents in and out of these pieces of spin chains
by lattice vibrations. The first scenario naturally leads to a constant
mean free path, but a theoretical prediction for the more likely
second scenario is presently lacking.

Recently, several theoretical papers considered magnetothermal
corrections (MTC) to the spin thermal conductivity, which should appear in a
magnetic field due to a coupling of the heat and magnetization currents.
\cite{Louis03,Sakai05_MTE,Sakai03_XXZ,HeidrichMeisner05}  In this
case, $\kappa_s$ is given, instead of
Eq.~(\ref{eMFTKappa}), by\cite{HeidrichMeisner05,HeidrichMeisner05PhD}
\begin{equation}\label{eMFTKappaMTC}
\kappa_s =  \frac{Na}{\pi} \left(  L_2  - \frac{L_1^2}{L_0} \right). 
\end{equation}
The second term in parentheses represents the MTC.
However, in Ref.~\onlinecite{Shimshoni03} it was argued that MTC 
are absent in macroscopic samples of real materials where
the conservation of the total magnetization parallel to $B$ is broken
by spin-orbit coupling, prohibiting a piling up of magnetization.  In
Fig.~\ref{KHcalc}(b), we show $\kappa_s(B)$ calculated within the
relaxation time approximation for one temperature  without (solid
line) and with (dashed line) MTC.
According to these calculations, MTC should
completely destroy the plateau-like feature at $B_c$, obviously at variance
with the experiment.  Thus, our experiment provides strong evidence
against the existence of magnetothermal corrections to the spin
thermal conductivity in CuPzN.

The assumption of a constant mean free path agrees well with the
experiment at intermediate and high fields but predicts a
field-independent $\kappa_s$ at low fields for $T \ll J/k_B$.  
The experiment, however, shows a low-field minimum of $\kappa_s(B)$
located approximately at $\mu_B B \sim k_B T$ for low $T$, see
Fig.~\ref{PhDia}.  
This suggests strongly that the origin of this
anomalous behavior is related to the physics of the half-filled band
of Jordan-Wigner fermions.  Only excitations in a window of width $k_B
T$ around the $B=0$ Fermi surface are able to relax their momentum to
the lattice by Umklapp scattering. An explanation of the minimum
might be possible along the following lines. The spin excitations
which dominate the thermal transport have a typical energy of order $k_B T$,
those located directly at the Fermi surface with energy $0$ do not
contribute. Therefore, Umklapp scattering is most effective, if the
excitations with energy $k_B T$ (rather than $0$) have a momentum close to
$\pm \pi/(2 a)$, where Umklapp scattering is strongest, i.e. for
$\mu_B B \sim k_B T$. Indeed, if one assumes a scattering rate
with a peak at $k=\pm \pi/(2 a)$,  the
relaxation time approximation yields a minimum in $\kappa(B)$ at the right
position (not shown). An exponentially strong maximum in
$\kappa(B)$ has been theoretically predicted for clean spin chains
coupled to phonons in Ref.~\onlinecite{Shimshoni03} as a consequence of the
existence of certain approximate conservation laws. As our system is
disorder dominated, we do not believe that this mechanism is directly
applicable in the present situation. 
A minimum in $\kappa_s(B)$ was also predicted for a classical 1D Heisenberg model in Ref.~\onlinecite{Savin05} but with a different $T$ dependence.
A full theory of the low-field
minimum should include the interplay of Umklapp, impurity and phonon
scattering which is currently under investigation.

According to Eqs.~(\ref{eMFTKappa}, \ref{scaling}), at low $T$,
$\kappa_s(B_c,T) \simeq \kappa_s(0,T)/2$,
see Fig.~\ref{KHcalc}(b). 
Using this, we estimate the absolute values of the zero-field spin
contribution as $\kappa_s(0,T) \approx 2[\kappa(0,T)-\kappa(B_c,T)]$,
which is shown in Fig.~\ref{KT}. It is notable that $\kappa_s \propto
T^2$ implying a linear increase of the mean free path with $T$ as $l =
A T$, with $A = 1.0 \times 10^{-6} {\rm ~m/K}$ for $T \lesssim 1.5$~K.
This observation is in agreement with theoretical predictions for
weakly disordered spin chains, where the linear $T$ dependence arises
from the renormalization of the impurity potentials by Luttinger
liquid corrections, see e.g.  Refs.~\onlinecite{Eggert92,Rozhkov05}.

In summary, we have experimentally established the magnetic
contribution $\kappa_m$ to the thermal conductivity of the $S=1/2$
chain compound copper pyrazine dinitrate. At low temperatures, the
field dependence of $\kappa_m$ is characterized by two features, one
at high fields in the vicinity of $B_c$ and the other at low fields.
The low-field feature remains to be explained. The high-field
feature is associated to the spin thermal transport $\kappa_s$ with a
mean free path surprisingly weakly dependent on both field and
momentum. No magnetothermal corrections to $\kappa_s$ have been
identified in our experiment.

 \acknowledgments

We acknowledge useful discussions with P.~Jung and A.~K.~Kolezhuk. This work was
supported  by the DFG through SFB 608 and by the GIF.

\end{document}